\documentstyle[eqsecnum,aps]{revtex}

\makeatletter
\@addtoreset{equation}{section}
\makeatother

\newcommand{\bibi}{\bibitem}

\def\mlc{{m}}

\newcommand{\half}{\mbox{{\normalsize $\frac{1}{2}$}} }

\newcommand{\be}{\begin{equation}}
\newcommand{\ee}{\end{equation}}
\newcommand{\bea}{\begin{eqnarray}}
\newcommand{\eea}{\end{eqnarray}}
\newcommand{\eq}{\ref}
\newcommand{\beq}{\begin{equation}}
\newcommand{\eeq}{\end{equation}}

\newcommand{\lb}{\label}

\def \3{\ss}

\begin{document}
\draft
\preprint{TAUP-2454-97, HU-EP-97/60, Wash.~U.~HEP/97-62} 
\twocolumn[\hsize\textwidth\columnwidth\hsize\csname%
@twocolumnfalse\endcsname
\title{On Lattice Computations of $K^+ \rightarrow \pi^+ \pi^0$ Decay 
at $m_K =2m_\pi$
}
\author{Maarten F.L. Golterman and Ka Chun Leung}
\address{
Department of Physics, Washington University, \\
St. Louis, MO 63130, USA
}
\date{\today}
\maketitle
\begin{abstract}
We use one-loop chiral perturbation theory to compare potential lattice
computations of the $K^+\to\pi^+\pi^0$ decay amplitude at $m_K=2m_\pi$
with the experimental value.  We find that the combined one-loop
effect due to this unphysical pion to kaon mass ratio and typical
finite volume effects is still of order minus
20--30\%, and appears to dominate
the effects from quenching.
\end{abstract}
]                                                    %prep
\narrowtext
\section{Introduction}
\label{SEC1}

Recently, we have used chiral perturbation theory (ChPT) to one loop
in order to investigate three systematic effects which affect lattice
computations of the weak matrix element for $K^+\to\pi^+\pi^0$ decay:
quenching, finite-volume effects, and the use of unphysical values of
the quark masses and pion external momenta \cite{GL} (to which we will
refer as $I$).  (For an extension to the partially
quenched case, see ref. \cite{GL2}; for a recent lattice computation, see
\cite{jlqcd}; for other references,
see $I$.)  We considered the case of
a lattice computation (extrapolated to the continuum limit)
with three degenerate light quarks, and final-state
pions at rest.  This is unphysical, because SU(3)
flavor is broken in the real world, and because this choice of external
momenta does not conserve energy.

Here, we extend our results to the case $m_K=2m_\pi$, 
considering both the quenched and unquenched theories.  
This choice of masses conserves energy when all external mesons are 
at rest, and, since in reality $m_K\approx 3.6m_\pi$, it is 
closer to the real-world meson masses. 
One might therefore expect the
systematic errors for this choice of masses to be smaller than in the
energy-nonconserving case with degenerate masses.

The choice $m_K =2m_{\pi}$ was also advertized in ref.~\cite{Italian},
where it is used as one of the  ingredients
in an improvement program for lattice computations
of nonleptonic kaon decays with Wilson fermions.

Here, we restrict ourselves to a summary of our results, and a brief
discussion of differences with $I$; for other details we refer
to the extensive explanations contained in $I$.  We also discuss
similar systematic effects for $B_K$.
The notation is the same as that of $I$.
 
\section{Lattice Method}
\label{SEC2}

On the lattice, usually operators with
zero spatial momentum, 
$O(t)=\sum_{\vec x} O({\vec x},t)$, are used.
For $K^+ \rightarrow \pi^+ \pi^0$ decay on the lattice, one computes 
the time-correlation function 
\bea
C(t_2 ,t_1)&&\equiv \langle 0|\pi^+ (t_2) \pi^0 (t_2)O_4 (t_1)K^- (0)|0\rangle 
\\
&&\!\!\!\!\!\!\!\!\!\!\!\!\! 
\longrightarrow
\frac
{\langle 0|\pi^+ (0)\pi^0 (0)|\pi^+ \pi^0 \rangle\langle K^+ |K^- (0)|0 \rangle}
{\langle \pi^+ \pi^0 |\pi^+ \pi^0 \rangle \langle K^+|K^+ \rangle} \times
\nonumber \\ 
&&\ \ \ \langle \pi^+ \pi^0 |O_4 (0)|K^+ \rangle 
e^{-E_{2\pi} (t_2-t_1)} e^{-m_K t_1}, \lb{asy}
\eea
where $O_4$ is the $\Delta S=1$, $\Delta I=3/2$ part of the weak effective
hamiltonian~\cite{wise}.
Eq.~(\eq{asy}) shows the dominant term for large time separations, 
from which the desired $K^+$-decay matrix element 
$\langle \pi^+ \pi^0 |O_4 |K^+ \rangle$ 
can be extracted. This matrix element represents the decay 
process, confined to a finite lattice volume 
(we will restrict ourselves to a cubic volume with 
linear dimension $L$ and periodic boundary conditions),  
in which a kaon, at rest, decays into a state consisting of 
two pions at rest with the lowest energy $E_{2\pi}$ \cite{lues}. 
The case for which $m_K =2m_{\pi}$ is of special interest, since, in that
case, the matrix element corresponds to an energy-conserving 
process (in the infinite-volume limit), 
in contrast to, for instance, the mass-degenerate case  ($m_K =m_\pi$)  
in which energy  is injected through the weak operator $O_4$.

The aim of this paper is to  compare real-world physical 
quantities with those obtained from a hypothetical
lattice computation in full or quenched QCD with $m_K =2m_\pi$, using 
ChPT. 
Here, we will refer to the choice $m_K =2m_\pi$, as ``lattice masses." 
Also, we will only consider the case with unbroken isospin,
$m_u =m_d \not= m_s$.

\section{Analytic results for $\mlc_K =2\mlc_\pi$ from one-loop ChPT}
\label{SEC3}

To $O(p^4)$ in (quenched) ChPT, the one-loop diagrams (a) to (d) of Fig. 1 in 
$I$, along with the relevant wavefunction renormalizations, as well as 
the tree-level contributions of $O(p^4)$ weak operators~\cite{kambor},  
have to be evaluated in order to obtain the $K^+$-decay matrix element. 
In $I$, where the mass-degenerate case was considered, 
special care  was taken to accommodate the 
kinematic situation in which energy is not conserved. 
In contrast, for $m_K =2m_\pi$ and in infinite volume, 
standard Feynman diagram techniques can be used.  
In a finite volume, it was shown in $I$ that (power-like) 
finite-volume corrections come exclusively 
from diagram (b) of Fig. 1, in which the final-state pions from the weak decay 
of the kaon rescatter.  We have derived the finite-volume corrections for  
$m_K \not= m_\pi$, and, in particular, for  $m_K =2m_\pi$ in the same way. 

The weak decay operator $O_4$ does not couple directly to the singlet meson
$\eta_0$. However, for $m_u =m_d \not= m_s$, mixing between $\eta_8$ and $\eta_0$  
occurs. As a result, the $\eta$ two-point function inherents 
the ``double pole" of the quenched theory~\cite{bg1}, introducing 
dependence on the parameters $\delta$ and $\alpha$ through $\eta$ loops
($\delta=m_0^2/(24\pi^2f_\pi^2)$, with $m_0$ the singlet part of the 
$\eta'$ mass), which is not present in the mass-degenerate case. 
It turns out that 
these contributions are finite (independent of the cutoff),  
and, therefore, unambiguously predicted in quenched ChPT.

The one-loop result for the unquenched case follows directly from 
Eqs.~(43,44) (re-expressing the $1/f_\pi^3$ in terms of the bare decay
constant $f$ using Eq.~(7) of $I$) by substituting $m_K=2m_\pi$.  
In the quenched case (which in
$I$ was only considered for $m_K=m_\pi$), we obtain for $m_K=2m_\pi$, 
ignoring the contributions from $O(p^4)$ operators,
\bea
&&\langle \pi^+ \pi^0 |O_4 |K^+ \rangle^q= 
{9i\alpha^q_{\scriptscriptstyle 27}
\over \sqrt{2}f_q^3}m^2_K \times \lb{qO4}\\ 
&&\Biggl( 1+\frac{m^2_K}{(4\pi f_q)^2} 
\left[ -2\log{\frac{m^2_K} {\Lambda_q^2}} + c_K
+\frac{1}{4}G(m_\pi L)\right] \nonumber\\
&&\phantom{
\Biggl( 1+\frac{m^2_K}{(4\pi f_q)^2}\Bigl[
}
+\delta c_\delta
+\alpha c_\alpha\frac{m^2_K}{(4\pi f_q)^2} 
 \Biggr) 
\ ,\nonumber
\eea
with 
\bea
c_K&=&\frac{7}{12}\log{4}
+\frac{1}{12} \log{7}
-\frac{1}{2}\sqrt{3} \arctan{\frac{\sqrt{3}}{5}}\;, \lb{cs}\\
c_\delta&=&-\frac{5}{6}-2\log{4}+\frac{17}{9}\log{7}\;, \nonumber\\
c_\alpha&=&\frac{14}{9}+2\log{4} -\frac{431}{208} \log{7}
-\frac{2}{\sqrt{3}} \arctan{\frac{\sqrt{3}}{5}}\;, \nonumber
\eea
and
\be
G(x)=\frac{17.827}{x} +\frac{10\pi^2}{x^3}\;.   \lb{fv}
\ee
The super/subscript $q$ denotes quenched quantities: $f_q$ is the bare
decay constant of the quenched theory, {\it etc}. The first factor
of Eq.~(\eq{qO4}) is the tree-level result; the factor in parentheses
gives the one-loop correction factor.

Before substituting $m_K =2m_\pi$ in order to obtain Eq.~(\eq{qO4}), 
one finds nonanalytic contributions coming from poles at $m_\pi^2$, $m_K^2$ 
and $2m^2_K -m^2_\pi$ (which is the mass of a pure $s\bar s$ 
meson). These nonanalytic contributions have the values $c_K$, 
$c_\delta$ and $c_\alpha$ after substituting $m_K=2m_\pi$.
The reason that we do not display the more general result here is
that it is different depending on whether energy is conserved, or
all external mesons are at rest.  Only at $m_K=2m_\pi$ do both more
general expressions agree with each other.

The finite-volume correction (term proportional to $G(m_\pi L)$ in
Eq.~(\eq{qO4})) has the same form in the
full and quenched theories, since it  comes only
from the pion-pion rescattering diagram, which
has no internal quark loops.
Note that the function $G(x)$ is different from the one that appears
in the mass-degenerate case, {\it cf.}~Eq.~(84) in $I$.  

For $m_K\ne m_\pi$, $O(p^4)$ weak operators lead to contributions
to the one-loop correction factor of the form
$(Am^2_\pi +Bm_\pi m_K +Cm^2_K )/(4\pi f_q)^2$. In the 
mass-degenerate limit, this leaves a dependence on the linear
combination $A+B+C$; whereas for $m_K =2m_\pi$, the 
linear combination is $A+2B+4C$. In Eq.~(\eq{qO4}), we 
have set $A+2B+4C=0$. Needless to say, for a comparison of 
complete $O(p^4)$ results between different masses,  
information will be  needed about the values of the coefficients 
$A$, $B$ and $C$. 

\section{Comparisons}
\label{SEC4}
 
In this section, we will compare a hypothetical
lattice computation, in a finite volume and with $m_K=2m_\pi$, of the 
$K^+\to\pi^+\pi^0$ 
matrix element and $B_K$ with the real world.  We will denote 
the lattice meson masses by $m_{K,latt}$ and $m_{\pi,latt}=
\half m_{K,latt}$, and the real-world
masses by $m_\pi$ and $m_K$, with $m_\pi=136$~MeV and $m_K=496$~MeV.
We will use
$f_{\pi,latt}=f_\pi=132$ MeV ($f_\pi$ will only appear in one-loop
corrections).  The mass of the $\eta$ is always determined from the
tree-level relation $m^2_\eta =\frac{4}{3}m^2_K -\frac{1}{3}m^2_\pi$.

Little is known about the values of $O(p^4)$-operator 
coefficients. Therefore, as in $I$, we 
will ignore contributions from $O(p^4)$ operators, 
and use the values $770$~MeV and $1$~GeV for the cutoffs 
$\Lambda$ and $\Lambda_q$ of the full and 
quenched theories respectively, 
and take the spread as an indication of the uncertainty from the
lack of information about the  $O(p^4)$ constants. In subsection \ref{A}, 
we will consider the case of a full-QCD lattice computation, and
in subsection \ref{B} the quenched case.

\subsection{Full QCD at $m_K =2m_\pi$}
\label{A}

In the full theory, the $K^+\to\pi^+\pi^0$ matrix element 
for the real world (subscript $phys$) 
and on the lattice (subscript $latt$)
can be related to one loop using 
Eqs.~(43,44,7) of $I$ (which was derived for arbitrary $m_K$ and
$m_\pi$). For the real world, one evaluates  the result for
$\langle \pi^+ \pi^0 |O_4 |K^+\rangle$ 
at the physical values of $m_K$ and $m_\pi$; 
whereas on the lattice, one chooses 
$m_{\pi,latt}=\half m_{K,latt}$ and therefore
$m^2_\eta=5m^2_{K,latt}/4$.  We obtain 
\be
\langle \pi^+ \pi^0 |O_4 |K^+\rangle^f_{phys} = 
X\; 
\frac{4}{3} \frac{m^2_K -m^2_\pi}{m^2_{K,latt}}\;\langle\pi^+\pi^0|O_4  
|K^+\rangle^f_{latt}\ ,\lb{fmr}
\ee
where the superscript $f$ denotes the full theory, and where   
\be
X=\frac{1+U_{phys}} {1+{U_{latt}}+\frac{m^2_{K,latt}}
{4(4\pi f_\pi)^2}G(m_{K,latt}L/2)}  
\lb{oneloopX}
\ee
is the one-loop correction to the 
tree-level ``conversion factor" $4(m^2_K -m^2_\pi )/(3m^2_{K,latt})$ 
($U$ denotes the one-loop correction term ({\it cf.} Eq~(87) of $I$) 
in infinite volume, {\it i.e.} with $G=0$).
We will restrict ourselves to 
the case where the lattice kaon mass is the same as 
the physical kaon mass, {\it i.e.} $m_{K,latt} =m_K$, 
which will presumably be accessible in future lattice computations.
It is straightforward to consider other examples.

The tree-level conversion factor is unambiguous \cite{bs}, and, for
our example, equal to 1.23.  We will therefore concentrate on the
one-loop factor $X$.
 From Eq.~(87) of $I$, $U_{phys}=0.0888$ and $-0.0146$ for
$\Lambda=1$~GeV and $770$~MeV respectively (we will 
ignore the imaginary part of the matrix element, 
since it does not contribute to the magnitude of 
the amplitude to order $p^4$). 
For $m_{K,latt} =m_K$, 
we obtain $U_{latt}=0.328$ and $0.147$ 
for $\Lambda=1$~GeV and $770$~MeV respectively. 

On the lattice, when $L$ is such that $m_K L =6$ or $8$, 
the relative one-loop contribution from the finite volume correction
$\frac{m^2_K}{4(4\pi f_\pi)^2}G(m_K L/2)$ is $0.215$ 
or $0.134$, respectively. 
Since, as explained before, we have omitted {\it different} 
linear combinations of $O(p^4)$ coefficients in the real-world 
and lattice cases respectively, 
we will vary the values of the cutoff in $U_{phys}$ and $U_{latt}$ 
independently. 

We list in Table \ref{tb:TABLE1} the values of $X$ for 
four combinations of the values $1$~GeV or $770$~MeV for the cutoff, and for 
volumes such that $m_K L=6$, $m_K L=8$ and $m_K L=\infty$. 
We take the  spread of the factor $X$ due to changes in 
the values of the cutoff as a systematic error, 
which, from Table \ref{tb:TABLE1}, 
is around  $15-20$\%.  

The one-loop expression for $B_K$ in the full theory is given by
Eq.~(36) of $I$.  For physical masses, $B^{f,\ phys}_K /B^f =1.72$ or
$1.42$ for  $\Lambda=1$~GeV or $770$~MeV, respectively;
whereas for $m_{K,latt}=2m_{\pi,latt}=m_K$,
$B^{f,latt}_K /B^f =1.73$ or
$1.44$, for the same values of the cutoff.
If we compare at the same value of the cutoff, we see that
the real-world and lattice values differ by about 1\%,
in contrast to the $K^+$-decay matrix element.
However, if we compare at different values of the cutoff, again as an
estimate of the error introduced by ignoring $O(p^4)$ coefficients,
we see that the real-world and lattice values may differ by as much
as 20\%.

Following $I$, we can examine the ratio
\be
{\cal R}=\left[ f_K\; \langle \pi^+ \pi^0 |O_4 |K^+ \rangle /\;
\langle {\overline K}^0 |O'|K^0 \rangle \right]^f_{latt}\;,\lb{R}
\ee

\begin{table}
\begin{center}
\begin{tabular}{c|cccc}
$m_K L$&$X^{(1)}_{(1)}$&$X^{(0.77)}_{(0.77)}$
&$X^{(1)}_{(0.77)}$&$X^{(0.77)}_{(1)}$ \\ 
\hline\\
$6$&0.71&0.72&0.80&0.64 \\ 
$8$&0.74&0.77&0.85&0.67 \\ 
$\infty$&0.82&
0.86&0.95&0.74 \\  
\end{tabular}
\end{center}
\caption{
The factor $X$ for different values of $m_K L$ and different combinations
of values of the cutoff $\Lambda$ (super/subscripts $(1)$ and $(0.77)$
denote cutoff used in the numerator/denominator of Eq.~(\eq{oneloopX})).
}
\vspace{0.2cm}
\label{tb:TABLE1}
\end{table}

\noindent
which is independent of the
$O(p^2)$-operator coefficient $\alpha_{\scriptscriptstyle 27}$.  
The tree-level value for the above ratio is $9i/8\sqrt{2}$.
At one loop, we find, for $m_{K,latt} =2m_{\pi,latt}=m_K$,
corrections of $-53$\%, $-61$\% and $-74$\% for $\Lambda=1$~GeV
and $m_\pi L=6$, $8$ and $\infty$, respectively (for
$\Lambda=770$~MeV, the corrections are $-29$\%, $-37$\% and $-50$\%).

\subsection{Quenched QCD at $m_K =2m_\pi$}
\label{B}

We now compare real-world quantities and quantities as would be
obtained from a quenched lattice computation with
$m_K =2m_\pi$. 
We get, from Eq.~(\eq{qO4}),  
\bea
\langle \pi^+ \pi^0 |&&O_4 |K^+ \rangle^q_{unphys}= 
{9i\alpha^q_{\scriptscriptstyle 27}
\over \sqrt{2}f_q^3}m^2_{K,latt}\times  \lb{2qO4} \\
&&\left( 1+U^q_{latt}+{m^2_{K,latt} \over 4(4\pi f_q)^2} 
G(m_{K,latt} L/2)\right)\ , \nonumber 
\eea
with
\bea
U^q_{latt}=
&&{m_{K,latt} \over (4\pi f_q)^2} \left(-2\log{m^2_{K,latt} \over 
\Lambda_q^2} + 0.6820
\right) \lb{V} \\ 
&&\phantom{11}
+0.0697\delta +0.0603\alpha {m^2_{K,latt} \over (4\pi f_q)^2}\ . 
\nonumber
\eea

In analogy with Eq.~(\eq{fmr}), we  relate the real-world $K^+$ decay 
matrix element to that of the quenched lattice theory, obtaining 
\bea
\langle \pi^+ \pi^0 |&&O_4|K^+\rangle^f_{phys} =
Y{\alpha_{\scriptscriptstyle{27}}
\over \alpha^q_{\scriptscriptstyle{27}}}\left( f_q \over f\right)^3 \times
\lb{qmr} \\ 
&& 
{4\over 3} {m^2_K -m^2_\pi \over m^2_{K,latt}}\; \langle \pi^+ \pi^0 |O_4  
|K^+\rangle^q_{latt}\ ,\nonumber
\eea
where
\be
Y={1+U_{phys}
\over 1+U^q_{latt}+{m^2_{K,latt} \over 4(4\pi f_q)^2} G(m_{K,latt} L/2)} \lb{Y}
\ee
is again the one-loop correction to the tree-level conversion factor.

Again, we will restrict ourselves 
to the case $m_{K,latt}=2m_{\pi,latt}=m_K$, 
and take $f_q=f_\pi$. First,  
we estimate the importance of the $\delta$- and $\alpha$-dependent terms in 
$U^q_{latt}$, {\it cf.} 
Eq.~(\eq{V}). The $\delta$- and $\alpha$-independent part of $U^q_{latt}$ 
is equal to $0.69$ and $0.59$ for 
$\Lambda_q =1$ GeV and $\Lambda_q=770$ MeV respectively. The 
value for $\delta$ is estimated  
to be less than around $0.2$ \cite{delta}. $\alpha$ is only poorly known,
but is unlikely to be larger than one in magnitude \cite{delta}.  
Therefore, the total contribution of the  
$\delta$ and $\alpha$ terms is less than about ten percent of 
the $\delta$- and $\alpha$-independent contribution.  
We will omit the $\delta$ and $\alpha$ terms in Table \ref{tb:TABLE2} below.
Table \ref{tb:TABLE2} gives the values of $Y$ for 
the four combinations of the values $1$~GeV or $770$~MeV for the cutoffs,
and for volumes such that $m_K L=6$, $m_K L=8$ and $m_K L=\infty$. 
These values of $Y$ deviate substantially from the tree-level
value $Y=1$, like in the examples with $m_K=m_\pi$ considered in $I$.
We see that the  spread in the values  of the factor $Y$
is $15$\% or less.  

For the quenched one-loop expression for $B_K$,
we refer to Eq.~(37) of $I$. 
For $m_{K,latt}=2m_{\pi,latt}=m_K$,
\bea
\frac{B^{q,latt}_K}{B^q}&=&1-0.14\delta\lb{Bq}\\
&&\phantom{1}+\cases{0.691+0.050\alpha, &$\Lambda_q =1$ GeV\cr
        0.358+0.085\alpha, &$\Lambda_q =770$ MeV\cr}. \nonumber
\eea
The $\delta$ and $\alpha$ terms are again relatively
small.  If we compare with the values of
$B^{f,phys}_K/B^f$ (see previous subsection), we see that again
real-world and quenched lattice values differ only by a few percent
if we compare at the same value of the cutoffs.  At different
values of the cutoffs, they may again differ by as much as 20\%.

In the quenched case, the ratio $\cal R$ defined in Eq.~(\eq{R}) is,
for $m_{K,latt}=2m_{\pi,latt}=m_K$ and $f_q =f_\pi$,
\bea
{\cal R}_q&=&\left[f_K\; \frac{\langle \pi^+ \pi^0 |O_4 |K^+ \rangle}
{\langle {\overline K}^0 |O'|K^0 \rangle}\right]^q_{latt}=
{9i\over 8\sqrt{2}}\times \lb{qR} \\
&&\Biggl( 1+0.056\delta +0.0224G(m_K L/2) \nonumber \\
&&\phantom{\Biggl(}
+\cases{-0.380-0.058\alpha, &$\Lambda_q =1$ GeV\cr
               -0.140-0.093\alpha, &$\Lambda_q =770$ MeV\cr}\Biggr)\;.
\nonumber
\eea
We see that the quenched values of $\cal R$ are closer to the tree-level
value than the unquenched values, for the same choice of meson masses and
volumes.

\section{Conclusion}
\label{SEC5}

Our main results are the ratios $X$ and $Y$ in
Eqs.~(\eq{oneloopX},\eq{Y}), which give the one-loop
ChPT correction to the conversion factor between a lattice
computation of the $K^+\to\pi+\pi^0$ amplitude with $m_K=2m_\pi$
and the experimental
value, for unquenched and quenched QCD, respectively.  These ratios
give an estimate of the systematic effect due to finite
volume, unphysical quark% 

\begin{table}
\begin{center}
\begin{tabular}{c|cccc}
 &$Y^{(1)}_{(1)}$&$Y^{(0.77)}_{(0.77)}$&$Y^{(1)}_{(0.77)}$&$Y^{(0.77)}_{(1)}$ \\ 
\hline\\
$m_K L=6$&0.71&0.69&0.76&0.65 \\ 
$m_K L=8$&0.75&0.73&0.81&0.68 \\ 
$\infty\  {\rm volume}$&0.83&
0.81&0.89&0.75 \\  
\end{tabular}
\end{center}
\caption{
The factor $Y$ for different values of $m_K L$ and different combinations
of values of the cutoffs $\Lambda$ and $\Lambda_q$
(super/subscripts $(1)$ and $(0.77)$
denote cutoff used in the numerator/denominator of Eq.~(\eq{Y})).
}
\vspace{0.2cm}
\label{tb:TABLE2}
\end{table}

\noindent
masses, and quenching ($Y$)
for this matrix element.  (The tree-level conversion factor corrects only
for unphysical masses.) For more discussion on the reliability
of such estimates, and other systematic errors, see $I$.

In Tables \ref{tb:TABLE1} and \ref{tb:TABLE2}, we give numerical examples,
illustrating these results for a lattice kaon mass 
equal to the experimental value, $m_K=496$~MeV.  The four different 
values on each line represent four combinations of cutoffs
we chose in evaluating the ratios
$X$ and $Y$, and we take the spread as an indication of the uncertainties
introduced by the lack of information on $O(p^4)$ constants.  We should
emphasize that ChPT does not give us any information about the ratio
$(\alpha_{\scriptscriptstyle 27}f_q^3)
/(\alpha^q_{\scriptscriptstyle 27}f^3)$
in Eq.~(\eq{qmr}).

{}From the Tables, we conclude that even in the ``more physical" case
$m_K=2m_\pi$, the one-loop systematic effect may still be as large as minus
20--30\%.
Comparing Tables \ref{tb:TABLE1} and \ref{tb:TABLE2}, we see that
quenching seems to have only a minor effect.  The large deviations
from one of the ratios $X$ and $Y$ are caused by the sensitivity of
the matrix element to the ratio $m_\pi/m_K$ and finite-volume effects.

This situation is different from that of $B_K$.  Here, if we compare
values of $B_K$ between the real world and the lattice at the same
value of the cutoff, the difference is only a few percent, both 
quenched and unquenched.  However, comparing at different 
values, again in order to get an idea of the effect of the
$O(p^4)$ constants,
indicates that also here systematic effects can be as large as 20\%.

\medskip
We thank Tanmoy Bhattacharya and Rajan Gupta 
for discussions.  This work is supported in part by the US
Department of Energy (MG by the Outstanding Junior Investigator program).

\end{document}